\documentclass[rnote]{aa}  

\usepackage{graphicx}
\usepackage{txfonts}
\usepackage{natbib} 
\bibpunct{(}{)}{;}{a}{}{,} 

\usepackage{verbatim}
\usepackage{url}

\begin{document}
   \title{Cumulative physical uncertainty in modern stellar models}

   \subtitle{II. The dependence on the
chemical composition}

   \author{G. Valle \inst{1}, M. Dell'Omodarme \inst{1}, P.G. Prada Moroni \inst{1,2}, S. Degl'Innocenti \inst{1,2}
          }

   \authorrunning{Valle, G. et al.}

   \institute{Dipartimento di Fisica ``Enrico Fermi'',
Universit\`a di Pisa, largo Pontecorvo 3, Pisa I-56127 Italy
\and
  INFN,
 Sezione di Pisa, Largo B. Pontecorvo 3, I-56127, Italy}

   \offprints{G. Valle, valle@df.unipi.it}

   \date{Received 21/01/2013; accepted 08/04/2013}

  \abstract
   {}
   {  
We extend our previous work on the
effects of the uncertainties on the main input physics for the evolution of
low-mass stars. 
We analyse the dependence of the cumulative physical uncertainty
affecting stellar tracks on the chemical composition.  
}
{  
We
calculated more than 6000 stellar tracks and isochrones, with metallicity
ranging from $Z$ = 0.0001 to 0.02, by changing the following physical inputs
within their current range of uncertainty: $^{1}$H(p,$\nu e^+$)$^{2}$H,
$^{14}$N$(p,\gamma)^{15}$O and triple-$\alpha$ reaction rates, radiative and
conductive opacities, neutrino energy losses, and microscopic diffusion
velocities.
The analysis was performed using a latin hypercube sampling design.  We examine
in a statistical way -- for different metallicities -- the dependence  on the
variation of the physical inputs of the turn-off (TO) luminosity, the
central hydrogen exhaustion time ($t_{\rm H}$), the luminosity and the 
helium core mass at the red-giant branch (RGB) tip, and the zero age horizontal branch (ZAHB) luminosity in the
RR Lyrae region.
}
  {
For the stellar tracks, an increase in the metallicity from $Z$ = 0.0001
to $Z$ = 0.02 produces a cumulative physical uncertainty error variation in TO
luminosity from 0.028 dex to 0.017 dex, while the global uncertainty on
$t_{\rm H}$ increases from 0.42 Gyr to 1.08 Gyr.  For the RGB tip, the
cumulative uncertainty on the luminosity is almost constant at 0.03 dex,
whereas the one the helium core mass decreases from 0.0055 $M_{\sun}$ to
0.0035 $M_{\sun}$.  The dependence of the ZAHB luminosity error is not
monotonic with $Z$, and it varies from a minimum of 0.036 dex at $Z$ = 0.0005
to a maximum of 0.047 dex at $Z$ = 0.0001.  Regarding stellar isochrones of 12
Gyr, the cumulative physical uncertainty on the predicted TO luminosity
and mass increases respectively from 0.012 dex to 0.014 dex and from 0.0136
$M_{\sun}$ to 0.0186 $M_{\sun}$. Consequently, from $Z$
= 0.0001 to $Z$ = 0.02 for ages typical of galactic globular clusters, the
uncertainty on the age inferred from the TO luminosity increases from 325 Myr
to 415 Myr.
 }
{}


   \maketitle

\section{Introduction}\label{sec:intro}

The present paper extends the analysis presented in \citet{incertezze1}
(hereafter PI) on the impact of the physical uncertainties affecting stellar
evolution models to a wide range of chemical compositions covering the typical
values for stellar populations in the Milky Way and in the near galaxies.  As
in the previous work, this paper is focused on the evolutionary characteristics
of ancient stellar populations, and thus the analysis is restricted to stellar
models of low-mass stars from the main sequence (MS) to the zero age
horizontal branch (ZAHB).

In PI we determined, for a fixed chemical composition (metallicity $Z$ =
0.006, and helium abundance $Y$ =
0.26), the range of variation in some important evolutionary features due to
the changes in the input physics within their uncertainty, pointing out the
effect of the different physical inputs at different stages of stellar
evolution.

The aim of this paper is to analyse how the results found in PI depend on the
assumed chemical composition.  Thus, we computed models covering a wide
metallicity range, i.e. from $Z$ = 0.0001 to 0.02.
We focus on the turn-off luminosity $L_{\rm BTO}$, the central hydrogen
exhaustion time $t_{\rm H}$, the luminosity $L_{\rm tip}$ and the helium-core
mass $M_{\rm c}^{\rm He}$ at the red giant branch (RGB) tip, and the ZAHB
luminosity $L_{\rm HB}$ in the RR Lyrae region at $\log T_{\rm eff}$ = 3.83.
As turn-off luminosity we adopted the luminosity of a point brighter and 100 K
lower than the turn-off (hereafter BTO), a technique similar to the one
proposed by \citet{Chaboyer1996} for isochrones in the (B-V, V) plane and
already adopted in PI.

Table \ref{table:inputfisici} lists the
physical input that was allowed to vary and the assumed uncertainty on each.
We refer to PI for a detailed discussion underlying their choice and the
determination of the uncertainty ranges.

\begin{table}[ht]
\centering
\caption{Physical inputs perturbed in the calculations and their assumed
  uncertainty. The abbreviations used in the
    following tables are defined in parentheses.} 
\label{table:inputfisici}
\centering
\begin{tabular}{llc}
  \hline\hline
 description & parameter & uncertainty \\
 \hline
  $^{1}$H(p,$\nu e^+$)$^{2}$H reaction rate (pp)      & $p_1$ & 3\%  \\
  $^{14}$N(p,$\gamma$)$^{15}$O reaction rate ($^{14}$N)     & $p_2$ & 10\% \\
  radiative opacity ($k_{\rm r}$)                   & $p_3$ & 5\%  \\
  microscopic diffusion velocities ($v_{\rm d}$)             & $p_4$ & 15\% \\
  triple-$\alpha$ reaction rate  (3$\alpha$)               & $p_5$ & 20\% \\
  neutrino emission rate  ($\nu$)                      & $p_6$ & 4\%  \\
  conductive opacity ($k_{\rm c}$)                  & $p_7$ & 5\%  \\
   \hline
\end{tabular}
\tablefoot{The table is taken from PI and is reported for the reader's
  convenience.} 
\end{table}

\section{Description of the technique}\label{sec:metodo}

We adopt a reference model of $M = 0.90$ $M_{\sun}$ to explore the cumulative
effect of the uncertainty on the main input physics, for six different
metallicity values, i.e. $Z$ = 0.0001, 0.0005, 0.001, 0.006, 0.01, and 0.02.

Determination of the initial helium abundances was done using a linear
helium-to-metal enrichment law: $Y=Y_p+\frac{\Delta Y}{\Delta Z}Z$, with
cosmological $^4$He abundance $Y_p=0.2485$
\citep{cyburt04,steigman06,peimbert07a,peimbert07b}.  We adopted $\Delta
Y$/$\Delta Z=2$, as the typical value for this quantity, which is still
affected by several 
important sources of uncertainty \citep{pagel98,jimenez03,flynn04,gennaro10}.
Thus, we computed models by adopting the following couples of initial
metallicity 
and helium abundance ($Z$, $Y$): (0.0001, 0.249), (0.0005, 0.25), (0.001,
0.25), (0.006, 0.26), (0.01, 0.268), and (0.02, 0.288).

The adopted stellar evolutionary code, FRANEC, is the same as used in PI and for
the construction of the Pisa Stellar Evolution Data
Base\footnote{\url{http://astro.df.unipi.it/stellar-models/}} for low-mass
stars \citep{database2012}. We refer to these papers and to \citet{cefeidi}
for a detailed description of the input on the stellar evolutionary code and
of the ZAHB construction technique.

We briefly recall the technique followed in PI, i.e. a systematic variation
in the input physics on a fixed grid within their current uncertainty.  For
every physical input listed in Table \ref{table:inputfisici}, we adopted a
three-value multiplier $p_i$ with value 1.00 for the unperturbed case and
values $1.00 \pm \Delta p_i$ for perturbed cases ($\Delta p_i$ is the
uncertainty listed in Table \ref{table:inputfisici}). For each stellar track
calculation, a set of multiplier values (i.e. $p_1, \ldots, p_7$) is fixed.
Then, to cover the whole parameter space and explore all the possible
interactions among input physics, calculations of stellar tracks must be
performed for a full crossing, i.e. each parameter value $p_i$ crossed with
all the values of the other parameters $p_j$, with $j \neq i$.

However, as shown in PI, the interactions among
the physical inputs are negligible for such small perturbations.  This makes
possible to reduce the 
computational burden by avoiding computation of the whole set of $3^7 = 2187$
tracks 
with the same mass, chemical composition, and $\alpha_{\rm ml}$ for each
metallicity.  We select randomly -- for each $Z$ -- a subset of $n$ = 162
models to be computed using a latin hypercube sampling design: it is an
extension of the latin square to higher dimensions and it has optimal property
in 
reducing the variance of the estimators obtained from the linear models
\citep{Stein1987}.  The sample size was chosen to balance the expected
  run time with good precision in the estimation of the effects of the
  input physics.  In particular, we evaluated the impact of a sample size
  reduction on the significance of the estimated coefficients reported in
  PI. For this purpose, we considered that the estimated errors of the
  effects (see Sect.~\ref{sec:statistic_track}) of the input physics -- 
  hence the confidence intervals width 
  of the effects -- scale approximately as the square root of the ratio of
  the original sample size (i.e. 2187) and the subsample size $n$.  

  Test
  simulations, performed by extracting different samples of different size $n$
  and 
  analysing the influence of the input physics in each sample, showed that the
  sample size $n$ = 162 is sufficient to obtain robust estimates of the input
  physics impacts.  The random selection was performed using the R library
lhs \citep{LHS}, resulting in a total sample of 972 stellar tracks.

\section{Statistical analysis of physical uncertainty}
\label{sec:statistic_track}

Relying on such a large set of perturbed stellar models, we statistically
analysed the effects of varying the physical inputs listed in Table
\ref{table:inputfisici} on the selected evolutionary features. For each Z
value, we built linear regression models by extracting the values of the
dependent variable under study (i.e. $L_{\rm BTO}$, $t_{\rm H}$, $L_{\rm tip}$,
$M_{\rm c}^{\rm He}$, and $L_{\rm HB}$) from the stellar tracks and regressing
them against the covariates, i.e. the values of the parameter $p_i$.  Thus we
stratified the sample according to $Z$ values and we treated strata
separately.

An alternative approach would require a global fit of the data, modelling the
dependence on $Z$ and $Y$ by inserting of interaction terms into the
statistical model; however, direct computations showed that the models of
these dependencies are quite cumbersome and require a function with several
powers of log-metallicity. We therefore prefer not to rely on these complex
models and 
to put our trust in the simpler ones obtained by stratification.

The statistical models include only the predictors but no interactions among
them, since they can be safely neglected, as we showed in PI.  For each studied
evolutionary feature, we included as covariates in the model only the physical
input that can have an actual influence: for $L_{\rm BTO}$ and $t_{\rm H}$
the first four parameters of Table \ref{table:inputfisici}, and  all the
parameters for the
quantities related to later evolutionary stages.  The
models were fitted to the data with a least-squares method using the software
R 2.15.1 \citep{R}.

For each $Z$ value, for $\log L_{\rm BTO}$ and $t_{\rm H}$ the regression
models are
\begin{equation}
\log L_{\rm BTO}, t_{\rm H}  = \beta_0 + \sum_{i=1}^4 \beta_i \; p_i 
\label{eq:BTO-tHc}
\end{equation}
where $\beta_0, \ldots, \beta_4$ were the regression coefficients to be
estimated by the fit and 
$p_1, \ldots, p_4$ the perturbation multipliers defined in
Table~\ref{table:inputfisici}.

In the cases of $\log L_{\rm tip}$, $\log L_{\rm HB}$, and $M_{\rm c}^{\rm He}$, 
the linear models for each $Z$ value are
\begin{equation}
\log L_{\rm tip}, \log L_{\rm HB}, M_{\rm c}^{\rm He}  = \beta_0 + 
\sum_{i=1}^7 \beta_i \; p_i
\label{eq:RGBHB}
\end{equation}
where we added the dependence on the triple-$\alpha$ reaction rate, the
neutrino 
emission rate, and the conductive opacity 
$k_{\rm c}$.

The cumulative effect of the physical input perturbation obtained
from the statistical models for each evolutionary feature is listed in
Table~\ref{table:tot-evol}.  There we report in
the second column the reference value of the studied quantity obtained from
the model with unperturbed physics, for each $Z$ value, in the third column
the cumulative 
statistical impact of the various physical inputs defined as $\sum_i \Delta
p_i \times |\beta_i|$.

\begin{table}[ht]
\centering
\caption{Cumulative impact of the uncertainty on the physical inputs on the
  considered evolutionary features. }
\label{table:tot-evol}
\begin{tabular}{lrr}
  \hline\hline
$Z$ & Reference & Impact \\ 
  \hline
\multicolumn{3}{c}{$\log L_{\rm BTO}$ (dex)}\\
\hline
  0.0001 & 0.749 & 0.028 \\ 
  0.0005 & 0.648 & 0.024 \\ 
  0.0010 & 0.584 & 0.023 \\ 
  0.0060 & 0.354 & 0.020 \\ 
  0.0100 & 0.270 & 0.019 \\ 
  0.0200 & 0.151 & 0.017 \\ 
   \hline
\multicolumn{3}{c}{$t_{\rm H}$ (Gyr)}\\
  \hline
0.0001 & 7.21 & 0.42    \\ 
  0.0005 & 7.38 & 0.45  \\ 
  0.0010 & 7.74 & 0.48  \\ 
  0.0060 & 10.54 & 0.71 \\ 
  0.0100 & 12.32 & 0.84 \\ 
  0.0200 & 15.23 & 1.08 \\ 
   \hline
\multicolumn{3}{c}{$\log L_{\rm tip}$ (dex)}\\
  \hline
  0.0001 & 3.257 & 0.032 \\ 
  0.0005 & 3.315 & 0.031 \\ 
  0.0010 & 3.342 & 0.031 \\ 
  0.0060 & 3.406 & 0.030 \\ 
  0.0100 & 3.419 & 0.029 \\ 
  0.0200 & 3.431 & 0.029 \\ 
   \hline
\multicolumn{3}{c}{$M_{\rm c}^{\rm He}$ ($M_{\sun}$)}\\
  \hline
0.0001 & 0.5023 & 0.0055   \\ 
  0.0005 & 0.4936 & 0.0053 \\ 
  0.0010 & 0.4909 & 0.0048 \\ 
  0.0060 & 0.4836 & 0.0041 \\ 
  0.0100 & 0.4807 & 0.0037 \\ 
  0.0200 & 0.4750 & 0.0035 \\ 
   \hline
\multicolumn{3}{c}{$\log L_{\rm HB}$ (dex) at $\log T_{\rm
    eff}$ = 3.83}\\
  \hline
0.0001 & 1.792 & 0.047   \\ 
  0.0005 & 1.701 & 0.036 \\ 
  0.0010 & 1.671 & 0.037 \\ 
  0.0060 & 1.565 & 0.041 \\ 
  0.0100 & 1.522 & 0.043 \\ 
  0.0200 & 1.465 & 0.044 \\ 
   \hline
\end{tabular}
\tablefoot{First column: the metallicity of the
  stellar model; second column: the reference value for the model with
  unperturbed physical input; third column: the sum of the impacts
  of the analytical fit.}
\end{table}

\begin{figure*}
\centering
\includegraphics[width=5.6cm,angle=-90]{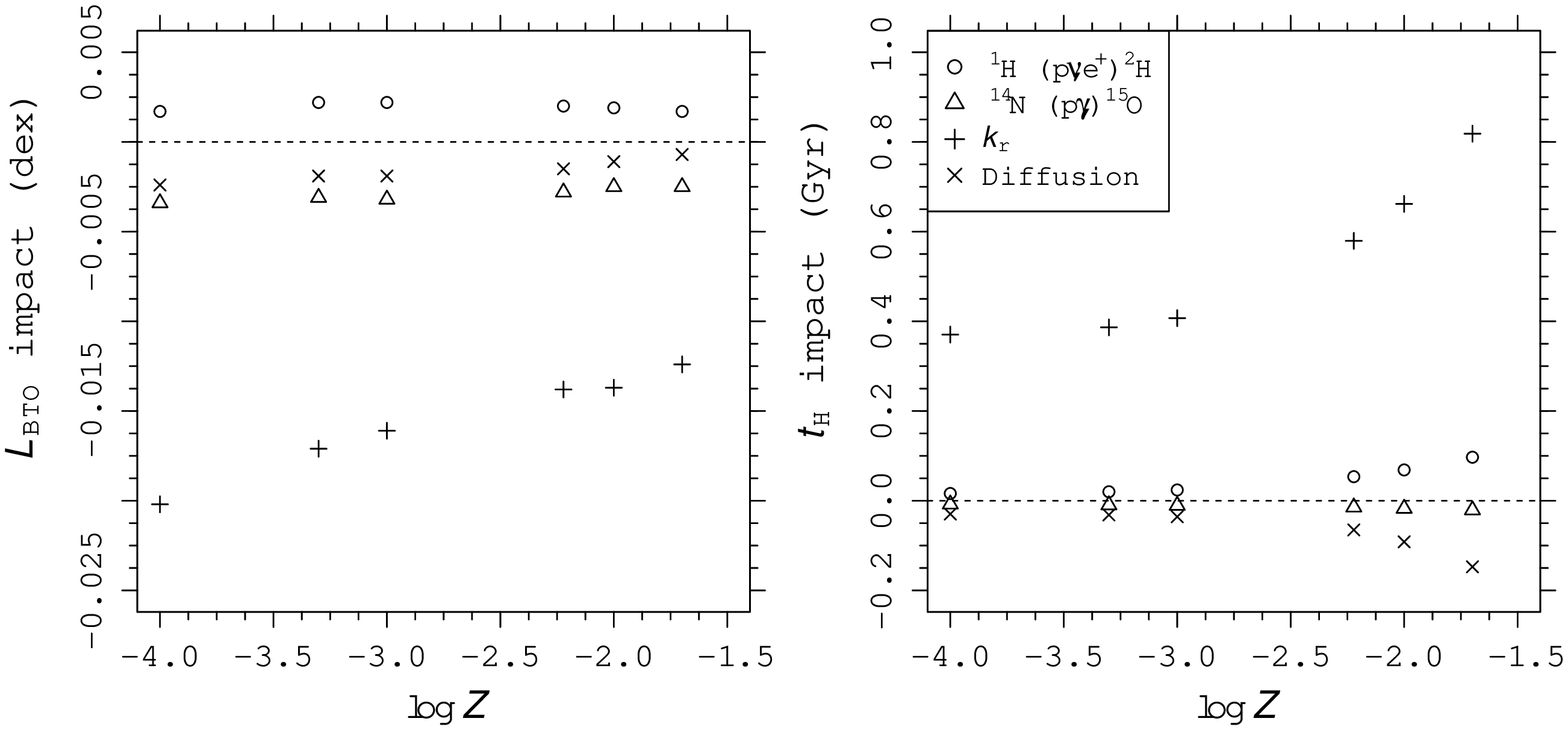}\\ 
\includegraphics[width=5.6cm,angle=-90]{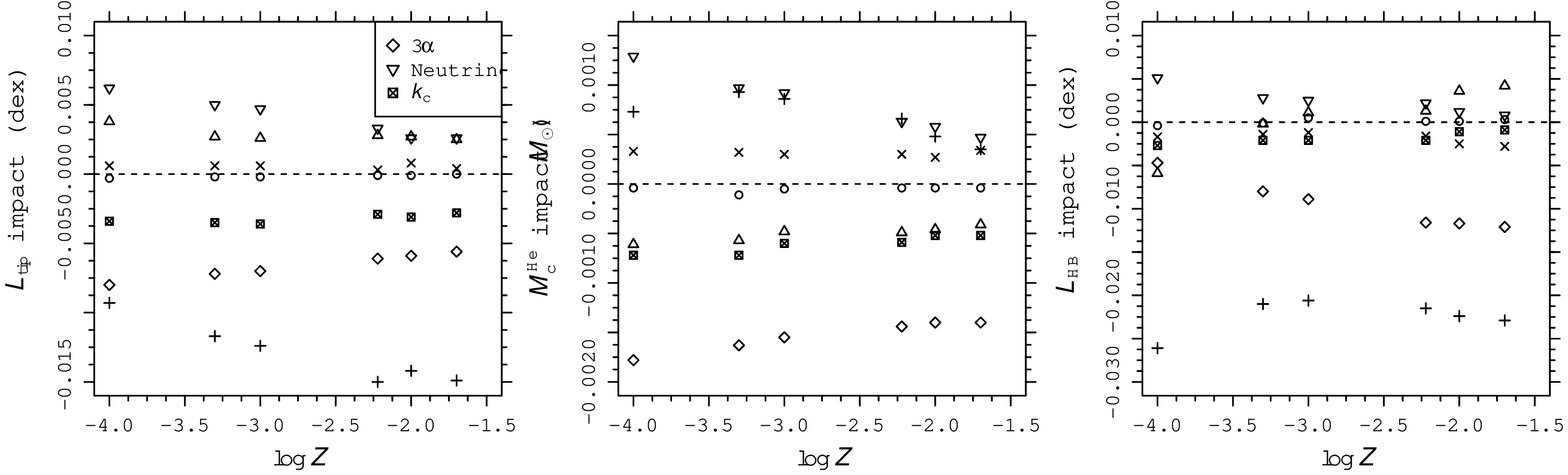}
\caption{Physical
impacts -- as in Tables~\ref{table:z-BTO}-\ref{table:z-HB} -- of the increment
$\Delta p_i$ of each multiplier $p_i$ for the 
considered physical inputs at different $\log Z$ values.}
\label{fig:impact}
\end{figure*}

>From Table~\ref{table:tot-evol} some general trends are apparent. In the
  case of BTO log-luminosity, the total impact decreases with metallicity from
  $Z$ = 0.0001 to 0.02 of about 40\%.  For central hydrogen exhaustion time,
  we note that the total impact grows with the metallicity with a rate
  slightly higher than the one of the increase in the reference time.  In the
  case of RGB tip, we observe that the total impact on the
  luminosity is nearly constant at
  about 0.03 dex, whereas on the helium core mass it decreases from 0.0055
  $M_{\sun}$ at $Z$ = 0.0001 to 0.0035 
  $M_{\sun}$ at $Z$ = 0.02.  In the case of ZAHB luminosity, the effect of
  the chemical composition on the cumulative physical uncertainty is not
  monotonic. In this case the total impact shows an initially high value at $Z$
  = 0.0001, a sudden drop at $Z$ = 0.0005, and a subsequent increase with
  the metallicity.

\onltab{
\begin{table}[ht]
\centering
\caption{Estimated coefficients for the linear model in Eq.~\ref{eq:BTO-tHc} of BTO luminosity
  (dex), for different $Z$ values.} 
\label{table:z-BTO}
\begin{tabular}{lrrrr}
  \hline\hline
 & Estimate & Std. Error & Impact & Importance \\ 
 & (dex)    & (dex)      & (dex)  & \\
  \hline
\multicolumn{5}{c}{$Z$ = 0.0001}\\
  \hline
$\beta_0$             & 1.1455 & 0.0026 & &  \\ 
$\beta_1$ (pp)        & 0.0581 & 0.0020 & 0.0017 & 0.086 \\ 
$\beta_2$ ($^{14}$N)   & -0.0338 & 0.0006 & -0.0034 & 0.167 \\ 
$\beta_3$ ($k_{\rm r}$) & -0.4050 & 0.0012 & -0.0202 & 1.000 \\ 
$\beta_4$ ($v_{\rm d}$) & -0.0160 & 0.0004 & -0.0024 & 0.119 \\ 
  \hline
\multicolumn{5}{c}{$Z$ = 0.0005}\\
  \hline
$\beta_0$ & 0.9593 & 0.0027 & & \\ 
$\beta_1$ (pp) & 0.0730 & 0.0021 & 0.0022 & 0.128 \\ 
$\beta_2$ ($^{14}$N) & -0.0312 & 0.0006 & -0.0031 & 0.183 \\ 
$\beta_3$ ($k_{\rm r}$) & -0.3411 & 0.0012 & -0.0171 & 1.000 \\ 
$\beta_4$ ($v_{\rm d}$) & -0.0124 & 0.0004 & -0.0019 & 0.109 \\ 
  \hline
\multicolumn{5}{c}{$Z$ = 0.001}\\
  \hline
$\beta_0$ & 0.8775 & 0.0026 & & \\ 
$\beta_1$ (pp) & 0.0735 & 0.0020 & 0.0022 & 0.137 \\ 
$\beta_2$ ($^{14}$N) & -0.0315 & 0.0006 & -0.0032 & 0.195 \\ 
$\beta_3$ ($k_{\rm r}$) & -0.3229 & 0.0012 & -0.0161 & 1.000 \\ 
$\beta_4$ ($v_{\rm d}$) & -0.0124 & 0.0004 & -0.0019 & 0.115 \\ 
  \hline
\multicolumn{5}{c}{$Z$ = 0.006}\\
  \hline
$\beta_0$ & 0.6031 & 0.0028 & &\\ 
$\beta_1$ (pp) & 0.0650 & 0.0021 & 0.0020 & 0.141 \\ 
$\beta_2$ ($^{14}$N) & -0.0278 & 0.0006 & -0.0028 & 0.202 \\ 
$\beta_3$ ($k_{\rm r}$) & -0.2760 & 0.0012 & -0.0138 & 1.000 \\ 
$\beta_4$ ($v_{\rm d}$) & -0.0098 & 0.0004 & -0.0015 & 0.106 \\ 
  \hline
\multicolumn{5}{c}{$Z$ = 0.01}\\
  \hline
$\beta_0$ & 0.5144 & 0.0030 & & \\ 
$\beta_1$ (pp) & 0.0620 & 0.0023 & 0.0019 & 0.136 \\ 
$\beta_2$ ($^{14}$N) & -0.0250 & 0.0007 & -0.0025 & 0.183 \\ 
$\beta_3$ ($k_{\rm r}$) & -0.2737 & 0.0014 & -0.0137 & 1.000 \\ 
$\beta_4$ ($v_{\rm d}$) & -0.0074 & 0.0005 & -0.0011 & 0.081 \\ 
  \hline
\multicolumn{5}{c}{$Z$ = 0.02}\\
  \hline
$\beta_0$ & 0.3715 & 0.0025 & & \\ 
$\beta_1$ (pp) & 0.0571 & 0.0019 & 0.0017 & 0.138 \\ 
$\beta_2$ ($^{14}$N) & -0.0248 & 0.0006 & -0.0025 & 0.200 \\ 
$\beta_3$ ($k_{\rm r}$) & -0.2484 & 0.0011 & -0.0124 & 1.000 \\ 
$\beta_4$ ($v_{\rm d}$) & -0.0044 & 0.0004 & -0.0007 & 0.053 \\ 
   \hline
\end{tabular}
\tablefoot{In the
first two columns: least-squares 
estimates of the regression coefficients and their errors; third column:
physical impact of the variation $\Delta p_i$ of the various inputs; last
column: relative importance of the impact with respect to the most important
one.}
\end{table}
}

\onltab{
\begin{table}[ht]
\centering
\caption{Estimated coefficients for the linear model in
  Eq.~\ref{eq:BTO-tHc} of the central hydrogen exhaustion time (Gyr),
  for different $Z$ 
  values. The column 
legend is the same as in Table~\ref{table:z-BTO}.}
\label{table:z-tHc}
\begin{tabular}{lrrrr}
\hline\hline
 & Estimate & Std. Error & Impact & Importance \\ 
 & (Gyr)    & (Gyr)      & (Gyr)  & \\
  \hline
\multicolumn{5}{c}{$Z$ = 0.0001}\\
  \hline
$\beta_0$ & -0.4596 & 0.0115 & & \\ 
$\beta_1$ (pp) & 0.5368 & 0.0089 & 0.016 & 0.043 \\ 
$\beta_2$ ($^{14}$N) & -0.0776 & 0.0027 & -0.008 & 0.021 \\ 
$\beta_3$ ($k_{\rm r}$) & 7.4100 & 0.0052 & 0.370 & 1.000 \\ 
$\beta_4$ ($v_{\rm d}$) & -0.1982 & 0.0017 & -0.030 & 0.080 \\ 
  \hline
\multicolumn{5}{c}{$Z$ = 0.0005}\\
  \hline
$\beta_0$ & -0.7055 & 0.0105 & & \\ 
$\beta_1$ (pp) & 0.6674 & 0.0081 & 0.020 & 0.052 \\ 
$\beta_2$ ($^{14}$N) & -0.0965 & 0.0024 & -0.010 & 0.025 \\ 
$\beta_3$ ($k_{\rm r}$) & 7.7301 & 0.0047 & 0.387 & 1.000 \\ 
$\beta_4$ ($v_{\rm d}$) & -0.2115 & 0.0016 & -0.032 & 0.082 \\ 
  \hline
\multicolumn{5}{c}{$Z$ = 0.001}\\
  \hline
$\beta_0$ & -0.8548 & 0.0130 & & \\ 
$\beta_1$ (pp) & 0.8001 & 0.0101 & 0.024 & 0.059 \\ 
$\beta_2$ ($^{14}$N) & -0.1047 & 0.0030 & -0.011 & 0.026 \\ 
$\beta_3$ ($k_{\rm r}$) & 8.1423 & 0.0059 & 0.407 & 1.000 \\ 
$\beta_4$ ($v_{\rm d}$) & -0.2379 & 0.0020 & -0.036 & 0.088 \\ 
  \hline
\multicolumn{5}{c}{$Z$ = 0.006}\\
  \hline
$\beta_0$ & -2.2543 & 0.0306 & & \\ 
$\beta_1$ (pp) & 1.7872 & 0.0237 & 0.054 & 0.092 \\ 
$\beta_2$ ($^{14}$N) & -0.1437 & 0.0071 & -0.014 & 0.025 \\ 
$\beta_3$ ($k_{\rm r}$) & 11.5876 & 0.0139 & 0.579 & 1.000 \\ 
$\beta_4$ ($v_{\rm d}$) & -0.4341 & 0.0046 & -0.065 & 0.112 \\ 
  \hline
\multicolumn{5}{c}{$Z$ = 0.01}\\
  \hline
$\beta_0$ & -2.4268 & 0.0421 & & \\ 
$\beta_1$ (pp) & 2.2898 & 0.0325 & 0.069 & 0.104 \\ 
$\beta_2$ ($^{14}$N) & -0.1721 & 0.0097 & -0.017 & 0.026 \\ 
$\beta_3$ ($k_{\rm r}$) & 13.2384 & 0.0191 & 0.662 & 1.000 \\ 
$\beta_4$ ($v_{\rm d}$) & -0.6116 & 0.0064 & -0.092 & 0.139 \\ 
  \hline
\multicolumn{5}{c}{$Z$ = 0.02}\\
  \hline
$\beta_0$ & -3.1791 & 0.0600 & & \\ 
$\beta_1$ (pp) & 3.2350 & 0.0464 & 0.097 & 0.119 \\ 
$\beta_2$ ($^{14}$N) & -0.2074 & 0.0139 & -0.021 & 0.025 \\ 
$\beta_3$ ($k_{\rm r}$) & 16.3678 & 0.0272 & 0.818 & 1.000 \\ 
$\beta_4$ ($v_{\rm d}$) & -0.9817 & 0.0091 & -0.147 & 0.180 \\ 
   \hline
\end{tabular}
\end{table}
}

\onltab{
\begin{table}[ht]
\centering
\caption{Estimated coefficients for the linear model in
  Eq.~\ref{eq:RGBHB} of RGB tip luminosity (dex), for different $Z$
  values. The column 
legend is the same as in Table~\ref{table:z-BTO}.}
\label{table:z-tip}
\begin{tabular}{lrrrr}
  \hline\hline
 & Estimate & Std. Error & Impact & Importance \\ 
 & (dex)    & (dex)      & (dex)  & \\
  \hline
\multicolumn{5}{c}{$Z$ = 0.0001}\\
  \hline
$\beta_0$ & 3.3628 & 0.0120 & & \\ 
$\beta_1$ (pp) & -0.0086 & 0.0077 & -0.0003 & 0.028 \\ 
$\beta_2$ ($^{14}$N) & 0.0377 & 0.0023 & 0.0038 & 0.405 \\ 
$\beta_3$ ($k_{\rm r}$) & -0.1858 & 0.0046 & -0.0093 & 1.000 \\ 
$\beta_4$ ($v_{\rm d}$) & 0.0041 & 0.0015 & 0.0006 & 0.066 \\ 
$\beta_5$ (3$\alpha$) & -0.0402 & 0.0011 & -0.0080 & 0.866 \\ 
$\beta_6$ ($\nu$) & 0.1552 & 0.0057 & 0.0062 & 0.668 \\ 
$\beta_7$ ($k_{\rm c}$) & -0.0676 & 0.0047 & -0.0034 & 0.364 \\ 
  \hline
\multicolumn{5}{c}{$Z$ = 0.0005}\\
  \hline
$\beta_0$ & 3.5049 & 0.0109 & & \\ 
$\beta_1$ (pp) & -0.0062 & 0.0070 & -0.0002 & 0.016 \\ 
$\beta_2$ ($^{14}$N) & 0.0271 & 0.0021 & 0.0027 & 0.231 \\ 
$\beta_3$ ($k_{\rm r}$) & -0.2349 & 0.0041 & -0.0117 & 1.000 \\ 
$\beta_4$ ($v_{\rm d}$) & 0.0037 & 0.0014 & 0.0006 & 0.047 \\ 
$\beta_5$ (3$\alpha$) & -0.0362 & 0.0010 & -0.0072 & 0.616 \\ 
$\beta_6$ ($\nu$) & 0.1258 & 0.0051 & 0.0050 & 0.428 \\ 
$\beta_7$ ($k_{\rm c}$) & -0.0693 & 0.0042 & -0.0035 & 0.295 \\ 
  \hline
\multicolumn{5}{c}{$Z$ = 0.001}\\
  \hline
$\beta_0$ & 3.5569 & 0.0101 & & \\ 
$\beta_1$ (pp) & -0.0069 & 0.0065 & -0.0002 & 0.017 \\ 
$\beta_2$ ($^{14}$N) & 0.0256 & 0.0019 & 0.0026 & 0.206 \\ 
$\beta_3$ ($k_{\rm r}$) & -0.2478 & 0.0039 & -0.0124 & 1.000 \\ 
$\beta_4$ ($v_{\rm d}$) & 0.0039 & 0.0013 & 0.0006 & 0.048 \\ 
$\beta_5$ (3$\alpha$) & -0.0351 & 0.0010 & -0.0070 & 0.566 \\ 
$\beta_6$ ($\nu$) & 0.1171 & 0.0048 & 0.0047 & 0.378 \\ 
$\beta_7$ ($k_{\rm c}$) & -0.0719 & 0.0039 & -0.0036 & 0.290 \\ 
  \hline
\multicolumn{5}{c}{$Z$ = 0.006}\\
  \hline
$\beta_0$ & 3.6840 & 0.0024 & & \\ 
$\beta_1$ (pp) & -0.0020 & 0.0015 & -0.0001 & 0.004 \\ 
$\beta_2$ ($^{14}$N) & 0.0277 & 0.0005 & 0.0028 & 0.185 \\ 
$\beta_3$ ($k_{\rm r}$) & -0.2997 & 0.0009 & -0.0150 & 1.000 \\ 
$\beta_4$ ($v_{\rm d}$) & 0.0021 & 0.0003 & 0.0003 & 0.021 \\ 
$\beta_5$ (3$\alpha$) & -0.0303 & 0.0002 & -0.0061 & 0.404 \\ 
$\beta_6$ ($\nu$) & 0.0816 & 0.0011 & 0.0033 & 0.218 \\ 
$\beta_7$ ($k_{\rm c}$) & -0.0574 & 0.0009 & -0.0029 & 0.192 \\ 
  \hline
\multicolumn{5}{c}{$Z$ = 0.01}\\
  \hline
$\beta_0$ & 3.7028 & 0.0132 & & \\ 
$\beta_1$ (pp) & -0.0048 & 0.0085 & -0.0001 & 0.010 \\ 
$\beta_2$ ($^{14}$N) & 0.0267 & 0.0025 & 0.0027 & 0.188 \\ 
$\beta_3$ ($k_{\rm r}$) & -0.2845 & 0.0050 & -0.0142 & 1.000 \\ 
$\beta_4$ ($v_{\rm d}$) & 0.0054 & 0.0017 & 0.0008 & 0.057 \\ 
$\beta_5$ (3$\alpha$) & -0.0296 & 0.0013 & -0.0059 & 0.416 \\ 
$\beta_6$ ($\nu$) & 0.0641 & 0.0062 & 0.0026 & 0.180 \\ 
$\beta_7$ ($k_{\rm c}$) & -0.0616 & 0.0051 & -0.0031 & 0.216 \\ 
  \hline
\multicolumn{5}{c}{$Z$ = 0.02}\\
  \hline
$\beta_0$ & 3.7215 & 0.0116 & & \\ 
$\beta_1$ (pp) & 0.0000 & 0.0075 & 0.0000 & 0.000 \\ 
$\beta_2$ ($^{14}$N) & 0.0249 & 0.0022 & 0.0025 & 0.167 \\ 
$\beta_3$ ($k_{\rm r}$) & -0.2988 & 0.0044 & -0.0149 & 1.000 \\ 
$\beta_4$ ($v_{\rm d}$) & 0.0030 & 0.0015 & 0.0004 & 0.030 \\ 
$\beta_5$ (3$\alpha$) & -0.0280 & 0.0011 & -0.0056 & 0.374 \\ 
$\beta_6$ ($\nu$) & 0.0642 & 0.0055 & 0.0026 & 0.172 \\ 
$\beta_7$ ($k_{\rm c}$) & -0.0553 & 0.0045 & -0.0028 & 0.185 \\ 
   \hline
\end{tabular}
\end{table}
}

\onltab{
\begin{table}[ht]
\centering
\caption{Estimated coefficients for the linear model in Eq.~\ref{eq:RGBHB} of
  helium core mass $M_{\rm c}^{\rm He}$ ($M_{\sun}$) for 
  different $Z$  values. The column 
legend is the same as in Table~\ref{table:z-BTO}.}
\label{table:z-McHe}
\begin{tabular}{lrrrr}
  \hline\hline
 & Estimate & Std. Error & Impact & Importance \\ 
 & ($M_{\sun}$)    & ($M_{\sun}$)      & ($M_{\sun}$)  & \\
  \hline
\multicolumn{5}{c}{$Z$ = 0.0001}\\
  \hline
$\beta_0$ & 0.4839 & 0.0042 & & \\ 
$\beta_1$ (pp) & -0.0013 & 0.0027 & -0.0000 & 0.023 \\ 
$\beta_2$ ($^{14}$N) & -0.0061 & 0.0008 & -0.0006 & 0.344 \\ 
$\beta_3$ ($k_{\rm r}$) & 0.0147 & 0.0016 & 0.0007 & 0.411 \\ 
$\beta_4$ ($v_{\rm d}$) & 0.0022 & 0.0005 & 0.0003 & 0.185 \\ 
$\beta_5$ (3$\alpha$) & -0.0089 & 0.0004 & -0.0018 & 1.000 \\ 
$\beta_6$ ($\nu$) & 0.0323 & 0.0020 & 0.0013 & 0.724 \\ 
$\beta_7$ ($k_{\rm c}$) & -0.0143 & 0.0016 & -0.0007 & 0.402 \\ 
  \hline
\multicolumn{5}{c}{$Z$ = 0.0005}\\
  \hline
$\beta_0$ & 0.4806 & 0.0021 & & \\ 
$\beta_1$ (pp) & -0.0037 & 0.0013 & -0.0001 & 0.069 \\ 
$\beta_2$ ($^{14}$N) & -0.0057 & 0.0004 & -0.0006 & 0.353 \\ 
$\beta_3$ ($k_{\rm r}$) & 0.0186 & 0.0008 & 0.0009 & 0.573 \\ 
$\beta_4$ ($v_{\rm d}$) & 0.0021 & 0.0003 & 0.0003 & 0.196 \\ 
$\beta_5$ (3$\alpha$) & -0.0081 & 0.0002 & -0.0016 & 1.000 \\ 
$\beta_6$ ($\nu$) & 0.0242 & 0.0010 & 0.0010 & 0.596 \\ 
$\beta_7$ ($k_{\rm c}$) & -0.0144 & 0.0008 & -0.0007 & 0.442 \\ 
  \hline
\multicolumn{5}{c}{$Z$ = 0.001}\\
  \hline
$\beta_0$ & 0.4749 & 0.0017 & & \\ 
$\beta_1$ (pp) & -0.0017 & 0.0011 & -0.0001 & 0.033 \\ 
$\beta_2$ ($^{14}$N) & -0.0048 & 0.0003 & -0.0005 & 0.312 \\ 
$\beta_3$ ($k_{\rm r}$) & 0.0172 & 0.0006 & 0.0009 & 0.555 \\ 
$\beta_4$ ($v_{\rm d}$) & 0.0020 & 0.0002 & 0.0003 & 0.194 \\ 
$\beta_5$ (3$\alpha$) & -0.0077 & 0.0002 & -0.0015 & 1.000 \\ 
$\beta_6$ ($\nu$) & 0.0230 & 0.0008 & 0.0009 & 0.596 \\ 
$\beta_7$ ($k_{\rm c}$) & -0.0120 & 0.0007 & -0.0006 & 0.388 \\ 
  \hline
\multicolumn{5}{c}{$Z$ = 0.006}\\
  \hline
$\beta_0$ & 0.4779 & 0.0004 & & \\ 
$\beta_1$ (pp) & -0.0013 & 0.0003 & -0.0000 & 0.027 \\ 
$\beta_2$ ($^{14}$N) & -0.0049 & 0.0001 & -0.0005 & 0.338 \\ 
$\beta_3$ ($k_{\rm r}$) & 0.0131 & 0.0001 & 0.0007 & 0.456 \\ 
$\beta_4$ ($v_{\rm d}$) & 0.0020 & 0.0001 & 0.0003 & 0.209 \\ 
$\beta_5$ (3$\alpha$)& -0.0072 & 0.0000 & -0.0014 & 1.000 \\ 
$\beta_6$ ($\nu$) & 0.0158 & 0.0002 & 0.0006 & 0.440 \\ 
$\beta_7$ ($k_{\rm c}$) & -0.0119 & 0.0002 & -0.0006 & 0.412 \\ 
  \hline
\multicolumn{5}{c}{$Z$ = 0.01}\\
  \hline
$\beta_0$ & 0.4782 & 0.0015 & & \\ 
$\beta_1$ (pp) & -0.0012 & 0.0010 & -0.0000 & 0.026 \\ 
$\beta_2$ ($^{14}$N) & -0.0046 & 0.0003 & -0.0005 & 0.331 \\ 
$\beta_3$ ($k_{\rm r}$) & 0.0095 & 0.0006 & 0.0005 & 0.340 \\ 
$\beta_4$ ($v_{\rm d}$) & 0.0018 & 0.0002 & 0.0003 & 0.195 \\ 
$\beta_5$ (3$\alpha$) & -0.0070 & 0.0001 & -0.0014 & 1.000 \\ 
$\beta_6$ ($\nu$) & 0.0145 & 0.0007 & 0.0006 & 0.413 \\ 
$\beta_7$ ($k_{\rm c}$) & -0.0104 & 0.0006 & -0.0005 & 0.372 \\ 
  \hline
\multicolumn{5}{c}{$Z$ = 0.02}\\
  \hline
$\beta_0$ & 0.4769 & 0.0015 & & \\ 
$\beta_1$ (pp) & -0.0015 & 0.0010 & -0.0000 & 0.031 \\ 
$\beta_2$ ($^{14}$N) & -0.0041 & 0.0003 & -0.0004 & 0.292 \\ 
$\beta_3$ ($k_{\rm r}$) & 0.0070 & 0.0006 & 0.0003 & 0.249 \\ 
$\beta_4$ ($v_{\rm d}$) & 0.0023 & 0.0002 & 0.0003 & 0.244 \\ 
$\beta_5$ (3$\alpha$) & -0.0070 & 0.0001 & -0.0014 & 1.000 \\ 
$\beta_6$ ($\nu$) & 0.0117 & 0.0007 & 0.0005 & 0.334 \\ 
$\beta_7$ ($k_{\rm c}$) & -0.0103 & 0.0006 & -0.0005 & 0.369 \\ 
   \hline
\end{tabular}
\end{table}
}

\onltab{
\begin{table}[ht]
\centering
\caption{Estimated coefficients for the linear model in Eq.~\ref{eq:RGBHB} of
  the ZAHB luminosity $\log L_{\rm HB}$ at $\log T_{\rm 
    eff}$ = 3.83 (dex) for
  different $Z$  values. The column 
legend is the same as in Table~\ref{table:z-BTO}.}
\label{table:z-HB}
\begin{tabular}{lrrrr}
  \hline\hline
 & Estimate & Std. Error & Impact & Importance \\ 
 & (dex)    & (dex)      & (dex)  & \\
  \hline
\multicolumn{5}{c}{$Z$ = 0.0001}\\
  \hline
$\beta_0$ & 2.3509 & 0.0168 & & \\ 
$\beta_1$ (pp) & -0.0150 & 0.0108 & -0.0004 & 0.017 \\ 
$\beta_2$ ($^{14}$N) & -0.0594 & 0.0032 & -0.0059 & 0.227 \\ 
$\beta_3$ ($k_{\rm r}$) & -0.5224 & 0.0064 & -0.0261 & 1.000 \\ 
$\beta_4$ ($v_{\rm d}$) & -0.0113 & 0.0021 & -0.0017 & 0.065 \\ 
$\beta_5$ (3$\alpha$) & -0.0237 & 0.0016 & -0.0047 & 0.182 \\ 
$\beta_6$ ($\nu$) & 0.1267 & 0.0079 & 0.0051 & 0.194 \\ 
$\beta_7$ ($k_{\rm c}$) & -0.0539 & 0.0065 & -0.0027 & 0.103 \\ 
  \hline
\multicolumn{5}{c}{$Z$ = 0.0005}\\
  \hline
$\beta_0$ & 2.1457 & 0.0091 & & \\ 
$\beta_1$ (pp) & -0.0019 & 0.0058 & -0.0001 & 0.003 \\ 
$\beta_2$ ($^{14}$N) & -0.0015 & 0.0017 & -0.0002 & 0.007 \\ 
$\beta_3$ ($k_{\rm r}$) & -0.4209 & 0.0035 & -0.0210 & 1.000 \\ 
$\beta_4$ ($v_{\rm d}$) & -0.0093 & 0.0012 & -0.0014 & 0.066 \\ 
$\beta_5$ (3$\alpha$) & -0.0399 & 0.0009 & -0.0080 & 0.380 \\ 
$\beta_6$ ($\nu$)& 0.0712 & 0.0043 & 0.0028 & 0.135 \\ 
$\beta_7$ ($k_{\rm c}$) & -0.0420 & 0.0035 & -0.0021 & 0.100 \\ 
  \hline
\multicolumn{5}{c}{$Z$ = 0.001}\\
  \hline
$\beta_0$ & 2.0878 & 0.0121 & & \\ 
$\beta_1$ (pp) & 0.0163 & 0.0078 & 0.0005 & 0.024 \\ 
$\beta_2$ ($^{14}$N) & 0.0106 & 0.0023 & 0.0011 & 0.052 \\ 
$\beta_3$ ($k_{\rm r}$) & -0.4126 & 0.0046 & -0.0206 & 1.000 \\ 
$\beta_4$ ($v_{\rm d}$) & -0.0078 & 0.0015 & -0.0012 & 0.057 \\ 
$\beta_5$ (3$\alpha$) & -0.0446 & 0.0012 & -0.0089 & 0.432 \\ 
$\beta_6$ ($\nu$) & 0.0623 & 0.0057 & 0.0025 & 0.121 \\ 
$\beta_7$ ($k_{\rm c}$) & -0.0415 & 0.0047 & -0.0021 & 0.101 \\ 
  \hline
\multicolumn{5}{c}{$Z$ = 0.006}\\
  \hline
$\beta_0$ & 2.0339 & 0.0045 & & \\ 
$\beta_1$ (pp) & 0.0040 & 0.0029 & 0.0001 & 0.006 \\ 
$\beta_2$ ($^{14}$N)& 0.0133 & 0.0009 & 0.0013 & 0.062 \\ 
$\beta_3$ ($k_{\rm r}$) & -0.4304 & 0.0017 & -0.0215 & 1.000 \\ 
$\beta_4$ ($v_{\rm d}$) & -0.0109 & 0.0006 & -0.0016 & 0.076 \\ 
$\beta_5$ (3$\alpha$) & -0.0579 & 0.0004 & -0.0116 & 0.538 \\ 
$\beta_6$ ($\nu$) & 0.0556 & 0.0021 & 0.0022 & 0.103 \\ 
$\beta_7$ ($k_{\rm c}$) & -0.0427 & 0.0017 & -0.0021 & 0.099 \\ 
  \hline
\multicolumn{5}{c}{$Z$ = 0.01}\\
  \hline
$\beta_0$ & 1.9961 & 0.0067 & & \\ 
$\beta_1$ (pp) & 0.0045 & 0.0043 & 0.0001 & 0.006 \\ 
$\beta_2$ ($^{14}$N) & 0.0362 & 0.0013 & 0.0036 & 0.161 \\ 
$\beta_3$ ($k_{\rm r}$) & -0.4489 & 0.0025 & -0.0224 & 1.000 \\ 
$\beta_4$ ($v_{\rm d}$) & -0.0168 & 0.0009 & -0.0025 & 0.112 \\ 
$\beta_5$ (3$\alpha$) & -0.0583 & 0.0006 & -0.0117 & 0.520 \\ 
$\beta_6$ ($\nu$)& 0.0309 & 0.0032 & 0.0012 & 0.055 \\ 
$\beta_7$ ($k_{\rm c}$)& -0.0215 & 0.0026 & -0.0011 & 0.048 \\ 
  \hline
\multicolumn{5}{c}{$Z$ = 0.02}\\
  \hline
$\beta_0$ & 1.9478 & 0.0056 & & \\ 
$\beta_1$ (pp) & 0.0084 & 0.0036 & 0.0003 & 0.011 \\ 
$\beta_2$ ($^{14}$N) & 0.0425 & 0.0011 & 0.0042 & 0.185 \\ 
$\beta_3$ ($k_{\rm r}$) & -0.4588 & 0.0021 & -0.0229 & 1.000 \\ 
$\beta_4$ ($v_{\rm d}$) & -0.0189 & 0.0007 & -0.0028 & 0.124 \\ 
$\beta_5$ (3$\alpha$) & -0.0603 & 0.0005 & -0.0121 & 0.526 \\ 
$\beta_6$ ($\nu$) & 0.0211 & 0.0027 & 0.0008 & 0.037 \\ 
$\beta_7$ ($k_{\rm c}$)& -0.0172 & 0.0022 & -0.0009 & 0.037 \\ 
   \hline
\end{tabular}
\end{table}
}

In Fig.~\ref{fig:impact} we display the results of the fits for the different
evolutionary quantities. This figure shows the physical impact of the
increment $\Delta p_i$ on each multiplier $p_i$ in dependence on $\log Z$.
The full results of the fits are reported in
Tables~\ref{table:z-BTO}-\ref{table:z-HB} (available online). These tables
list, for different values of $Z$, the least squares estimations of the
regression coefficients $\beta_i$, their standard error, the physical impact
of the increment of $\Delta p_i$ on each multiplier $p_i$, and the absolute
value of the relative importance of these impacts with respect to the most
important one.

In the case of $Z$ = 0.006 the values of the estimated coefficients are
different from the ones reported in PI, since the present estimates are
obtained from a subsample of the original dataset. As expected, in almost all
cases the present estimates and the ones of PI agree within their errors.

The figure and the tables show that the effect of perturbing the main physical
input within their current uncertainty on the studied stellar evolutionary
quantities changes in a rather complex way when varying the chemical
composition. In particular, the relative importance of the various input
physics in contributing to the global uncertainty changes with metallicity. As
a consequence, one should be careful to extrapolate the results we obtained in
PI for stellar models of $Z$ = 0.006 to significantly different metallicities.

In the case of BTO luminosity, from Fig.~\ref{fig:impact} and from the
last column of Table~\ref{table:z-BTO}, we note that the effect of the
variation in radiative opacity is dominant for each value of $Z$, while the
effect of the other physical inputs change mildly with metallicity. The second
most important input is the one due to the $^{14}$N(p,$\gamma$)$^{15}$O reaction
rate at each $Z$. The impact of the perturbations scales almost linearly with
$\log Z$.
 
For central hydrogen exhaustion time in Table~\ref{table:z-tHc}, it is apparent
that the importance of radiative opacity is dominant for each $Z$ and that
the contribution of microscopic diffusion and of $^{1}$H(p,$\nu e^+$)$^{2}$H
reaction rate increase with metallicity. At $Z$ = 0.02 they jointly account
for about 30\% of the impact of $k_{\rm r}$. Figure~\ref{fig:impact} shows that
the impact of the radiative opacity variation grows more than linearly with
$\log Z$.
 
In the case of RGB tip luminosity, radiative opacity is the most important
uncertainty source, but at lower $Z$ its importance is close to the one of
triple-$\alpha$ (87\%) and neutrino cooling (67\%); in contrast, for $Z$ =
0.02 the cumulative impact of all the other sources of uncertainty is about
80\% of the one of $k_{\rm r}$. Figure~\ref{fig:impact} shows that the effect is
due to the growth in the importance of the radiative opacities uncertainty
with $\log Z$, with a concurrent decrease in the importance of the other
sources of uncertainty.

For the helium core mass at RGB tip, the triple-$\alpha$ reaction rate is the
most important uncertainty source. A distinctive feature is the decrease in
the importance of neutrino cooling with increasing $Z$; while it accounts for
72\% of the impact of the triple-$\alpha$ at $Z$ = 0.0001, this
percentage decreases to 33\% at $Z$ = 0.02. We note also from
Fig.~\ref{fig:impact} a 
non-monotonic trend with $\log Z$ for the radiative opacities effect.

For ZAHB luminosity, the radiative opacity is the main source of
uncertainty at each metallicity. We observe from Fig.~\ref{fig:impact} an
increase in the importance of triple-$\alpha$ moving toward high $Z$. A
distinctive feature is the drop in the absolute value of the importance of
radiative opacity from $Z$ = 0.0001 to $Z$ = 0.0005, followed by a mild
increase with $\log Z$.  The trend in the cumulative uncertainty is further
complicated since the impact of $^{14}$N(p,$\gamma$)$^{15}$O, which is
negative at $Z$ = 0.0001, becomes positive at higher $Z$.  Moreover, each
physical input variation affects both the He-core mass (and thus indirectly
$L_{\rm HB}$) and $L_{\rm HB}$ directly.  These effects are the source of the
non-monotonic trend on the cumulative uncertainty shown in
Table~\ref{table:tot-evol}.  The precise understanding of the effect of each
physical input is made even more difficult by the  ZAHB
luminosity being evaluated at fixed $T_{\rm eff}$ for all $Z$.  This implies
that at low $Z$ the region is populated with higher masses with respect to the
ones at high $Z$. 

\section{Physical uncertainty on stellar isochrones}
\label{sec:iso}

The previous statistical analysis of the physical uncertainties affecting
stellar tracks can be extended to isochrones.  We focus here on isochrones
with ages in the range 8-14 Gyr, which is suitable for galactic globular
clusters.  As 
in PI, we are interested in studying the variation near the turn-off region;
so we performed calculations by varying only the physical input that can
influence the evolution until this phase, i.e. $p_1, \ldots, p_4$.  For each
$Z$ value, we selected stellar models from the full grid by varying $p_1,
\ldots, 
p_4$ (i.e. $3^4 = 81$ cases). Each metallicity grid contains several stellar
models with different masses, chosen to accurately reconstruct -- in the
desired age range -- the BTO region. The computed stellar models are listed
in Table~\ref{table:isocrone}.  A total of 5832 models and 3402 isochrones
were calculated.

\begin{figure*}
\centering
\includegraphics[width=5.6cm,angle=-90]{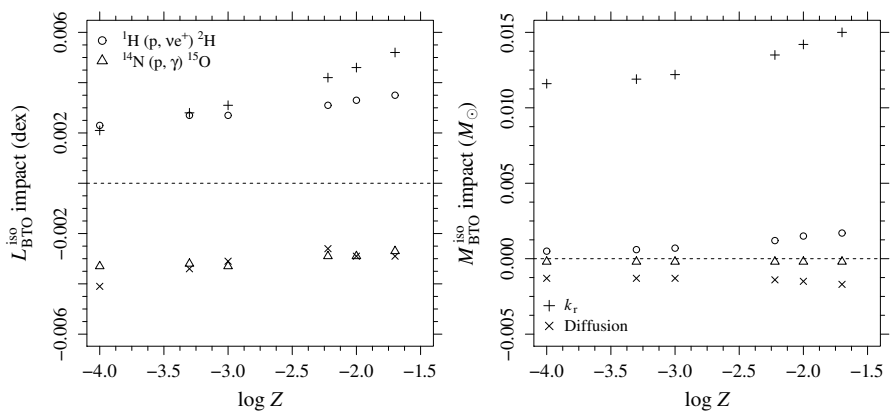}
\caption{Physical
impacts on isochrone BTO luminosity and mass at isochrone BTO
due to the increment $\Delta 
p_i$ of each multiplier $p_i$ for the 
considered physical inputs at different values of $\log Z$.}
\label{fig:impact-iso}
\end{figure*}

\onltab{ 
\begin{table}[ht]
\centering
\caption{Stellar models computed for isochrones construction at different
  $Z$ values.} 
\label{table:isocrone}
\begin{tabular}{ll}
  \hline\hline
$Z$ & Stellar models ($M_{\sun}$)\\ 
  \hline
  0.0001 &  0.50, 0.60, 0.70, 0.72, 0.75, 0.77, 0.80,\\
         &  0.82, 0.85, 0.87, 0.90, 0.95 \\ 
  0.0005 &  0.50, 0.60, 0.70, 0.72, 0.75, 0.77, 0.80,\\
         &  0.82, 0.85, 0.87, 0.90, 0.95 \\ 
  0.0010 &  0.50, 0.60, 0.70, 0.75, 0.77, 0.80, 0.82,\\
         &  0.85, 0.87, 0.90, 0.92, 0.95\\ 
  0.0060 &  0.50, 0.60, 0.70, 0.80, 0.85, 0.87, 0.90,\\
         &  0.92, 0.95, 1.00, 1.05, 1.10 \\ 
  0.0100 &  0.70, 0.80, 0.85, 0.87, 0.90, 0.92, 0.95,\\
         &  0.97, 1.00, 1.02, 1.05, 1.10 \\ 
  0.0200 &  0.70, 0.80, 0.85, 0.87, 0.90. 0.92, 0.95,\\
         &  0.97, 1.00, 1.02, 1.05, 1.10 \\ 
   \hline
\end{tabular}
\end{table}
}

Following the technique outlined in PI, we pool together the results obtained
from isochrones of different ages. For each $Z$ we first remove the trend due
to age by adapting an ANOVA model to the turn-off luminosity $\log L_{\rm
  BTO}^{\rm iso}$ and mass $M_{\rm BTO}^{\rm iso}$, with age as a categorical
predictor. This method increases the power of the analysis due to the larger
sample examined after pooling.  The residuals of these models are then
regressed against $p_1, \ldots, p_4$. Further details on the technique and
discussion of the underlying hypotheses can be found in PI.
 
The results about the total uncertainty on $\log L_{\rm BTO}^{\rm iso}$ and
$M_{\rm BTO}^{\rm iso}$, due to the simultaneous variation in all the
considered input physics, are shown in Table~\ref{table:totIso}.  As in
Sec.~\ref{sec:statistic_track} the total uncertainty is obtained by adding the
effect of each covariate.  The total uncertainty on the reconstructed $\log
L_{\rm BTO}^{\rm iso}$ and on the $M_{\rm BTO}^{\rm iso}$ shows a moderate
increase with $Z$. In the case of $\log L_{\rm BTO}^{\rm iso}$ the difference
between maximum and minimum uncertainty is of the order of 20\%, whereas for
$M_{\rm BTO}^{\rm iso}$ it is about 37\%.

\begin{table}[ht]
\centering
\caption{Cumulative impact on the uncertainty on isochrone BTO luminosity
  (dex) and on the 
  mass at isochrone BTO ($M_{\sun}$) for different metallicity values. }
\label{table:totIso}
\begin{tabular}{lrr}
  \hline\hline
& \multicolumn{2}{c}{Impact}\\
\hline
$Z$ & $\log L_{\rm BTO}^{\rm iso}$ (dex)& $M_{\rm BTO}^{\rm iso}$ ($M_{\sun}$)\\ 
  \hline
  0.0001 & 0.0118 & 0.0136\\ 
  0.0005 & 0.0121 & 0.0141\\ 
  0.0010 & 0.0122 & 0.0144\\ 
  0.0060 & 0.0129 & 0.0164\\ 
  0.0100 & 0.0138 & 0.0174\\ 
  0.0200 & 0.0143 & 0.0186\\ 
   \hline
\end{tabular}
\tablefoot{The results are obtained
  by pooling together isochrones of  
  ages in the range [8 - 14], see text for details.}
\end{table}

Figure~\ref{fig:impact-iso} and
Tables~\ref{table:z-isoLBTO}-\ref{table:z-isoMBTO} display the detailed
results of the fits for $\log L_{\rm BTO}^{\rm iso}$ and $M_{\rm BTO}^{\rm
  iso}$. The figure displays the physical impact of the increment of $\Delta
p_i$ of each multiplier $p_i$ in dependence on $\log Z$.  The tables report,
for different $Z$ values, the least squares estimations of the regression
coefficients $\beta_i$, their standard error, the physical impacts of the
increment of $\Delta p_i$ of each multiplier $p_i$, and the absolute value of
the relative
importance of these impact with respect to the most relevant one.

For $M_{\rm BTO}^{\rm iso}$ (Table~\ref{table:z-isoMBTO}) the relative
contribution of the physical sources of uncertainty is almost the same for all
the metallicity values; the only exception is the role of the p-p chain whose
relative contribution to the cumulative uncertainty slightly increases with
metallicity.  The radiative opacity dominantes over all the other
physical input.

In the case of $\log L_{\rm BTO}^{\rm iso}$ (Table \ref{table:z-isoLBTO})
the dependence on the chemical
composition shows much more complex behaviour. At $Z$ = 0.0001 the main
contribution to the uncertainty comes from the microscopic diffusion
velocities. At $Z$ = 0.001 the effect of microscopic diffusion velocities,
$^{14}$N(p,$\gamma$)$^{15}$O reaction rate, and radiative opacity are similar;
whereas at larger $Z$ the effect of $k_{\rm r}$ becomes dominant. As
shown 
in Table~\ref{table:totIso}, these variation nearly compensate
one another in the whole analysed range of metallicity, so that the cumulative
uncertainty on $\log L_{\rm BTO}^{\rm iso}$ is nearly constant and ranges from
0.012 dex to 0.014 dex.

\begin{table}[ht]
\centering
\caption{Effect of the various physical inputs on isochrones BTO
  luminosity (dex), for different $Z$ 
  values. The column 
legend is the same as in Table~\ref{table:z-BTO}.}
\label{table:z-isoLBTO}
\begin{tabular}{lrrrr}
  \hline\hline
 & Estimate & Std. Error & Impact & Importance \\ 
  \hline
\multicolumn{5}{c}{$Z$ = 0.0001}\\
  \hline
$\beta_0$ & -0.0591 & 0.0046 & & \\ 
$\beta_1$ (pp) & 0.0775 & 0.0038 & 0.0023 & 0.573 \\ 
$\beta_2$ ($^{14}$N) & -0.0332 & 0.0011 & -0.0033 & 0.817 \\ 
$\beta_3$ ($k_{\rm r}$) & 0.0418 & 0.0023 & 0.0021 & 0.515 \\ 
$\beta_4$ ($v_{\rm d}$) & -0.0271 & 0.0008 & -0.0041 & 1.000 \\ 
  \hline
\multicolumn{5}{c}{$Z$ = 0.0005}\\
  \hline
$\beta_0$ & -0.0919 & 0.0041 & & \\ 
$\beta_1$ (pp) & 0.0893 & 0.0034 & 0.0027 & 0.795 \\ 
$\beta_2$ ($^{14}$N) & -0.0319 & 0.0010 & -0.0032 & 0.945 \\ 
$\beta_3$ ($k_{\rm r}$) & 0.0569 & 0.0020 & 0.0028 & 0.844 \\ 
$\beta_4$ ($v_{\rm d}$) & -0.0225 & 0.0007 & -0.0034 & 1.000 \\ 
  \hline
\multicolumn{5}{c}{$Z$ = 0.001}\\
  \hline
$\beta_0$ & -0.0983 & 0.0038 & & \\ 
$\beta_1$ (pp) & 0.0896 & 0.0031 & 0.0027 & 0.826 \\ 
$\beta_2$ ($^{14}$N) & -0.0325 & 0.0009 & -0.0033 & 1.000 \\ 
$\beta_3$ ($k_{\rm r}$) & 0.0622 & 0.0018 & 0.0031 & 0.957 \\ 
$\beta_4$ ($v_{\rm d}$) & -0.0210 & 0.0006 & -0.0031 & 0.968 \\ 
  \hline
\multicolumn{5}{c}{$Z$ = 0.006}\\
  \hline
$\beta_0$ & -0.1400 & 0.0027 & & \\ 
$\beta_1$ (pp) & 0.1038 & 0.0022 & 0.0031 & 0.747 \\ 
$\beta_2$ ($^{14}$N) & -0.0295 & 0.0007 & -0.0029 & 0.707 \\ 
$\beta_3$ ($k_{\rm r}$) & 0.0834 & 0.0013 & 0.0042 & 1.000 \\ 
$\beta_4$ ($v_{\rm d}$) & -0.0177 & 0.0004 & -0.0026 & 0.635 \\ 
  \hline
\multicolumn{5}{c}{$Z$ = 0.01}\\
  \hline
$\beta_0$ & -0.1545 & 0.0031 & &  \\ 
$\beta_1$ (pp) & 0.1112 & 0.0026 & 0.0033 & 0.725 \\ 
$\beta_2$ ($^{14}$N) & -0.0293 & 0.0008 & -0.0029 & 0.637 \\ 
$\beta_3$ ($k_{\rm r}$) & 0.0920 & 0.0015 & 0.0046 & 1.000 \\ 
$\beta_4$ ($v_{\rm d}$) & -0.0195 & 0.0005 & -0.0029 & 0.635 \\ 
  \hline
\multicolumn{5}{c}{$Z$ = 0.02}\\
  \hline
$\beta_0$ & -0.1738 & 0.0031 & & \\ 
$\beta_1$ (pp) & 0.1164 & 0.0025 & 0.0035 & 0.673 \\ 
$\beta_2$ ($^{14}$N) & -0.0273 & 0.0008 & -0.0027 & 0.526 \\ 
$\beta_3$ ($k_{\rm r}$) & 0.1038 & 0.0015 & 0.0052 & 1.000 \\ 
$\beta_4$ ($v_{\rm d}$) & -0.0191 & 0.0005 & -0.0029 & 0.553 \\ 
   \hline
\end{tabular}
\end{table}

\begin{table}[ht]
\centering
\caption{Effect of the various physical inputs on the masses at isochrones BTO
  ($M_{\sun}$), for different $Z$ 
  values. The column 
legend is the same as in Table~\ref{table:z-BTO}.}
\label{table:z-isoMBTO}
\begin{tabular}{lrrrr}
  \hline
 & Estimate & Std. Error & Impact & Importance \\ 
  \hline
\multicolumn{5}{c}{$Z$ = 0.0001}\\
  \hline
$\beta_0$ & -0.2388 & 0.0019 & & \\ 
$\beta_1$ (pp) & 0.0165 & 0.0016 & 0.0005 & 0.042 \\ 
$\beta_2$ ($^{14}$N) & -0.0019 & 0.0005 & -0.0002 & 0.016 \\ 
$\beta_3$ ($k_{\rm r}$) & 0.2328 & 0.0009 & 0.0116 & 1.000 \\ 
$\beta_4$ ($v_{\rm d}$) & -0.0086 & 0.0003 & -0.0013 & 0.110 \\ 
  \hline
\multicolumn{5}{c}{$Z$ = 0.0005}\\
  \hline
$\beta_0$ & -0.2480 & 0.0015 & & \\ 
$\beta_1$ (pp) & 0.0207 & 0.0012 & 0.0006 & 0.052 \\ 
$\beta_2$ ($^{14}$N) & -0.0023 & 0.0004 & -0.0002 & 0.019 \\ 
$\beta_3$ ($k_{\rm r}$) & 0.2382 & 0.0007 & 0.0119 & 1.000 \\ 
$\beta_4$ ($v_{\rm d}$) & -0.0086 & 0.0002 & -0.0013 & 0.108 \\ 
  \hline
\multicolumn{5}{c}{$Z$ = 0.001}\\
  \hline
$\beta_0$ & -0.2558 & 0.0015 & & \\ 
$\beta_1$ (pp)& 0.0238 & 0.0012 & 0.0007 & 0.059 \\ 
$\beta_2$ ($^{14}$N) & -0.0025 & 0.0004 & -0.0002 & 0.020 \\ 
$\beta_3$ ($k_{\rm r}$) & 0.2432 & 0.0007 & 0.0122 & 1.000 \\ 
$\beta_4$ ($v_{\rm d}$) & -0.0087 & 0.0002 & -0.0013 & 0.107 \\ 
  \hline
\multicolumn{5}{c}{$Z$ = 0.006}\\
  \hline
$\beta_0$ & -0.2998 & 0.0021 & & \\ 
$\beta_1$ (pp) & 0.0413 & 0.0017 & 0.0012 & 0.092 \\ 
$\beta_2$ ($^{14}$N) & -0.0024 & 0.0005 & -0.0002 & 0.018 \\ 
$\beta_3$ ($k_{\rm r}$) & 0.2702 & 0.0010 & 0.0135 & 1.000 \\ 
$\beta_4$ ($v_{\rm d}$) & -0.0093 & 0.0003 & -0.0014 & 0.103 \\ 
  \hline
\multicolumn{5}{c}{$Z$ = 0.01}\\
  \hline
$\beta_0$ & -0.3200 & 0.0015 & & \\ 
$\beta_1$ (pp) & 0.0493 & 0.0012 & 0.0015 & 0.104 \\ 
$\beta_2$ ($^{14}$N) & -0.0023 & 0.0004 & -0.0002 & 0.016 \\ 
$\beta_3$ ($k_{\rm r}$) & 0.2833 & 0.0007 & 0.0142 & 1.000 \\ 
$\beta_4$ ($v_{\rm d}$) & -0.0103 & 0.0002 & -0.0015 & 0.109 \\ 
  \hline
\multicolumn{5}{c}{$Z$ = 0.02}\\
  \hline
$\beta_0$ & -0.3438 & 0.0018 & & \\ 
$\beta_1$ (pp) & 0.0576 & 0.0014 & 0.0017 & 0.115 \\ 
$\beta_2$ ($^{14}$N) & -0.0016 & 0.0004 & -0.0002 & 0.011 \\ 
$\beta_3$ ($k_{\rm r}$) & 0.2994 & 0.0009 & 0.0150 & 1.000 \\ 
$\beta_4$ ($v_{\rm d}$) & -0.0115 & 0.0003 & -0.0017 & 0.115 \\ 
   \hline
\end{tabular}
\end{table}

In Table~\ref{table:revage} we present the results of the analysis on the
uncertainty affecting the age inferred from BTO luminosity.  We
constructed -- for each $Z$ -- linear models linking $\log L_{\rm BTO}^{\rm
  iso}$ and the logarithm of the age. We ignore the dependence on the physical
inputs, confounding their effect with the residual standard error. The models
are then used to perform a reverse inference on the value of the age, given
$L_{\rm BTO}^{\rm iso}$. For each metallicity set, we fix the reference value
of $\log L_{\rm BTO}^{\rm iso}$ to the one reached at 12 Gyr. More details
about the technique are given in Appendix B of PI.  The results collected in
Table~\ref{table:revage} show that the uncertainty in the inferred age
increases by almost 30\% from $Z$ = 0.0001 to $Z$ = 0.02, ranging from 325
Myr to 415 Myr.
 
\begin{table}[ht]
\centering
\caption{Uncertainty on the age inferred from the BTO luminosity. 
  The uncertainty is evaluated around 12 Gyr.} 
\label{table:revage}
\begin{tabular}{lr}
  \hline\hline
$Z$ & Uncertainty (Gyr)\\ 
  \hline
  0.0001 & 0.325\\ 
  0.0005 & 0.345\\
  0.0010 & 0.355\\
  0.0060 & 0.345\\
  0.0100 & 0.365\\
  0.0200 & 0.415\\
   \hline
\end{tabular}
\end{table}

\section{Conclusions}
\label{sec:conclusion} 

In this paper we extended the work presented in \citet{incertezze1} that
analyses the cumulative uncertainty due to physical inputs in stellar
models of low-mass stars.  The aim was to analyse the influence on the results
of a variation in the assumed chemical composition. Thus we calculated stellar
models for a wide range of metallicity from $Z$ = 0.0001 to $Z$ = 0.02.

For the stellar tracks, the cumulative physical uncertainty in the analysed
quantities changes with the chemical composition.  For turn-off luminosity the
cumulative uncertainty decreases from 0.028 dex at $Z$ = 0.0001 to 0.017 dex
at $Z$ = 0.02. The global uncertainty on the central hydrogen exhaustion time
increases with $Z$ from 0.42 Gyr to 1.08 Gyr, with a rate that is slightly
higher than the increase in the reference time of hydrogen exhaustion at
any given $Z$. In the case of the RGB tip, the cumulative uncertainty on the
luminosity is almost constant at 0.03 dex, while on the helium core mass
it decreases from 0.0055 $M_{\sun}$ at $Z$ = 0.0001 to 0.0035 $M_{\sun}$ at $Z$ =
0.02.  The dependence on the metallicity of the error in the predicted ZAHB
luminosity is not monotonic. The total uncertainty shows a value of 0.047 dex
at $Z$ = 0.0001, a sudden drop to a value of 0.036 dex at $Z$ = 0.0005, and a
following increase with $Z$ up to a value of 0.044 dex at $Z$ = 0.02.
 
We confirm the results presented in PI; i.e., for all the analysed stellar
tracks features (except the He-core mass at the red giant branch tip), in the
full adopted range of chemical compositions, the effects of the uncertainty
on the radiative opacity tables is dominant. The uncertainty on the helium
core mass at RGB tip is instead mainly due to the variation in the
triple-$\alpha$ reaction rate. In conclusion, an increase in the precision of
the radiative opacity tables is mandatory to further improve stellar evolution
theoretical predictions.

For the stellar isochrones of 12 Gyr, the cumulative uncertainty on the
predicted turn-off luminosity and mass show a moderate increase with $Z$.
From $Z$ = 0.0001 to 0.02, the global physical uncertainty increases of the
order of 20\% for $\log L_{\rm BTO}^{\rm iso}$ and 37\% for  $M_{\rm
  BTO}^{\rm iso}$.  As a consequence, for ages in the range 8-14
Gyr, the uncertainty on the age inferred from the turn-off luminosity
increases by about 30\% from $Z$ = 0.0001 to $Z$ = 0.02, ranging from 325
Myr to 415 Myr.

\begin{acknowledgements}
We are grateful to our anonymous referee for pointing out a problem in the
draft version and for several suggestions that helped in clarifying and
improving the paper. This work has been supported by PRIN-INAF 2011 ({\em
  Tracing the 
  formation and evolution of the Galactic Halo with VST}, PI M. Marconi).
\end{acknowledgements}

\bibliographystyle{aa}
\bibliography{biblio}

\appendix
\section{On the EOS influence}
\label{app:eos}

As discussed in PI, the uncertainty on the equation-of-state cannot be
treated with the approach applied to the other physical input. 
To provide a rough estimate of the impact of
the present uncertainty on EOS, we computed the reference tracks --
i.e. the ones with unperturbed physics -- with two different and widely
adopted choices for
equation-of-state: OPAL EOS and FreeEOS \citep{irwin04}.

\begin{table}[ht]
\centering
\caption{Impact of equation of state change from OPAL to FreeEOS for the
 examined evolutionary features}
\label{table:eos}
\begin{tabular}{rrrrrr}
  \hline\hline
$Z$ & $\log L_{\rm BTO}$ & $t_{\rm H}$  & $\log L_{\rm tip}$ & $M_{\rm c}^{\rm
    He}$ & $\log L_{\rm HB}$ \\  
    & (dex) & (Gyr) & (dex) &($M_{\sun}$) & (dex)\\
  \hline
  0.0001 & 0.0017 & 0.053 & 0.0022 & -0.00095 & -0.0039 \\ 
         &  (6\%)   & (13\%)  & (7\%)    & (17\%)     & (8\%)\\
  0.0005 & 0.0025 & 0.049 & 0.0005 & -0.00117 & -0.0037 \\ 
         & (10\%)   & (11\%)  & (2\%)   & (22\%)     & (10\%)\\
  0.0010 & 0.0031 & 0.057 & -0.0006 & -0.00132 & -0.0046 \\ 
         & (13\%)   & (12\%)  & (2\%)   & (28\%)      & (12\%)\\
  0.0060 & 0.0028 & 0.096 & -0.0020 & -0.00149 & -0.0070 \\ 
         & (14\%)   & (13\%)  & (7\%)    & (36\%)   & (17\%)\\
  0.0100 & 0.0023 & 0.110 & -0.0027 & -0.00142 & -0.0030 \\ 
         & (12\%)   & (13\%)  & (9\%)     & (38\%)    & (7\%)\\
  0.0200 & 0.0031 & 0.192 & -0.0027 & -0.00115 & -0.0013 \\ 
         & (18\%)   & (18\%)  & (9\%)    & (33\%)     & (3\%)\\
   \hline
\end{tabular}
\tablefoot{In parenthesis: percentage influence of EOS change with respect to
  the values in the column ``Impact'' of Table~\ref{table:tot-evol}.}
\end{table}

Table~\ref{table:eos} reports -- for each evolutionary feature and for each
$Z$ -- the impact of the EOS change. A comparison of these values with the one
of the ``Impact'' column in Table~\ref{table:tot-evol} shows that the EOS
change accounts for about 10\% to 20\% of the total impact of all the other
physical input. The only exception is the helium core mass: in this case the
impact of the EOS variation accounts for 20\% to 40\% of the total impact of
the other 
physical input.

\end{document}